\documentclass[preprints,article,accept,pdftex,oneauthor]{Definitions/mdpi} 
\firstpage{1} 
\pubvolume{1}
\issuenum{1}
\articlenumber{0}
\pubyear{2023}
\copyrightyear{2023}
\datereceived{ } 
\daterevised{ } 
\dateaccepted{ } 
\datepublished{ } 
\hreflink{https://doi.org/} 

\usepackage[final]{changes}
\usepackage{nameref}

\usepackage{array}
\usepackage{amsmath,amssymb,latexsym}
\newcolumntype{L}[1]{>{\raggedright\let\newline\\\arraybackslash\hspace{0pt}}m{#1}}
\newcolumntype{C}[1]{>{\centering\let\newline\\\arraybackslash\hspace{0pt}}m{#1}}
\newcolumntype{R}[1]{>{\raggedleft\let\newline\\\arraybackslash\hspace{0pt}}m{#1}}
\usepackage{multirow}
\usepackage{multicol}
\usepackage{cprotect}
\usepackage{tabularx}

\newcommand{\wzurl}[1]{\href{#1}{\url{#1}}}

\Title{Scalable data concentrator with baseline interconnection network for triggerless data acquisition systems}

\TitleCitation{Scalable data concentrator with baseline interconnection network for triggerless data acquisition systems}

\Author{Wojciech M. Zabolotny $^{1,\dagger,\ddagger}$\orcidA{}}

\AuthorNames{Wojciech M. Zabolotny}

\AuthorCitation{Zabołotny, W.M.}

\address{%
$^{1}$ \quad Warsaw University of Technology, Faculty of Electronics and Information Technology, Institute of Electronic Systems, Nowowiejska 15/19, 00-650 Warszawa, Poland}	

\corres{Correspondence: ${*}$~wojciech.zabolotny@pw.edu.pl}

\abstract{Triggerless Data Acquisition Systems (DAQs) require transmitting the data stream from multiple links 
	to the processing node.
	The short input data words must be concentrated and packed into the longer 
	bit vectors the output interface (e.g., PCI Express) uses.
	In that process, the unneeded data must be eliminated, and a dense stream of
	useful DAQ data must be created. Additionally, the time order of the data should be preserved.
	This paper presents a new solution using the Baseline Network with Reversed Outputs (BNRO) for high-speed
	data routing. A thorough analysis of the network's operation enabled increased scalability compared to the
	previously published concentrator based on an 8x8 network.
 	The solution may be scaled by adding additional layers to the BNRO network while minimizing resource consumption.
	Simulations were done for 4 and 5 layers (16 and 32 inputs). The FPGA implementation and tests in the actual hardware have been successfully performed for 16 inputs. 
	The pipeline registers may be added in each layer independently, shortening the critical path and increasing the maximum acceptable clock frequency.}


\begin{document}


\section{Introduction}
The high-energy physics experiments generate vast volumes of data.
It is not possible to archive all that data for later offline processing.
Two approaches are used to solve that issue. 
In the triggered data acquisition systems, the incoming data is buffered until the ``level one (L1) trigger'' decides
whether it is useful. The data must be buffered until this decision is taken and delivered.
The L1 trigger decision must be elaborated quickly to keep the necessary capacity of data buffers reasonable, which usually requires high-performance FPGA chips.
When elaborating the local trigger decision and waiting for the L1 trigger decision, the data may be prepared for sending
to the data acquisition system (DAQ). 

However, using the triggered approach is not always possible. 
The papers~\cite{neufeld_future_2015, meschi_daq_2015} predicted the increasing role of triggerless DAQs eight years ago.
Some experiments already use or plan to use this approach: LHCb~\cite{colombo_lhcb_2018}, AMBER~\cite{zemko_triggerless_2022}, and  CBM~\cite{gao_throttling_2020}.
In that approach, the raw data from the detector are submitted only to elementary processing (e.g., zero suppression) and then delivered to the system responsible for identifying the interesting events -- the Event Selector.
The Event Selector may be built based on ATCA architecture~\cite{korcyl_modeling_2012} or as a tightly connected computer network~\cite{cuveland_first-level_2011,albrecht_event_2014}.
The second solution may be better scalable and upgradable, using the standard servers as the main component.
In that solution, it is still necessary to deliver the data from the Front-End Electronics (FEE) via the optical links to the computer being the entry nodes of the event building network.
That may be accomplished with specialized FPGA-based data concentrator boards connected to the PCIe bus
in the server~\cite{paramonov_felix_2021,cachemiche_pcie-based_2016} and equipped with multiple optical transceivers.

\deleted{
The data received from the detector is usually a stream of relatively short words. For example, in the CBM experiment~\cite{cbm_collaboration_technical_2023}, the STS detector readout chip - SMX2 delivers the hit data as 24-bit words~\cite{kasinski_protocol_2016}, which, after preprocessing and adding the source ID, will be extended to 32-bit words. The data are transmitted via GBT~\cite{baron_gbt_2009} links with a rate of ca. 10.67~Mwords/s~\cite{lehnert_gbt_2017}.
}
\added{The PCIe blocks in modern FPGA chips use a wide datapath to fully utilize the bandwidth offered by the PCIe interface at reasonable clock frequency. For example, the PCIe Gen3 block working with 8 lanes requires delivering 256 bits of data at a frequency of 250 MHz. For 16 lanes, the data should be delivered as 512-bit words at 250 MHz~\cite{url_xlx_pg195}.}

\added{The overall volume of data should be limited. Therefore, the data about the physics events recorded by detectors (so-called ``hit messages'') are transmitted as possibly short data words.
For example, in the CBM experiment~\cite{cbm_collaboration_technical_2023}, the STS detector readout chip - SMX2 delivers the hit data as 24-bit words~\cite{kasinski_protocol_2016}, which, after preprocessing and adding the source ID, will be extended to 32-bit words.
Similarly, the Detector Data Link (DDL) protocol, used in the ALICE experiment at CERN LHC~\cite{carena_ddl_2015}, is oriented toward transmitting 32-bit words.}

\added{Concatenating the shorter words received from links into the wider word used by PCIe seems simple. In the case of eight 32-bit links concentrated to a 256-bit PCIe word, a constant group of bits could be allocated for each link. However, such a solution would be inefficient.
The links do not transmit the hit data all the time. The data stream may contain additional words with time markers, control, status, and diagnostic information. If there is no data to be sent now, the ``empty'' words are transmitted.
Generally, the data may be divided into ``DAQ data'',
which should be transmitted to DAQ, and ``non-DAQ data'', which should be discarded.
If the PCIe word consists of bit groups permanently assigned to particular links, such discarded data would create ``holes'' in the data transmitted via PCIe and then stored in the DMA memory buffers in the Event Selector computer. That results in wasting the PCIe bandwidth and memory and may reduce the performance of data
processing.}

\deleted{
Each link provides a ``non-dense'' data stream, which should be filtered before the concentration.
The PCIe blocks in the FPGA chips require much wider data words. For example, the PCIe Gen3 block working with 8 lanes requires delivering 256 bits of data at a frequency of 250 MHz. For 16 lanes, the data should be delivered as 512-bit words at 250 MHz.
During the concentration, the short ``DAQ'' words arriving from input links should be packed into the wide words used by the PCIe blocks,
eliminating the ``non-DAQ'' words to avoid holes in the output data stream. Additionally, the data should be delivered to the PCIe in the same order as they were received from the links. 
The article~\cite{guminski_benes_2023} discusses various approaches to the data concentration. They may be divided into three methods.
The first method is based on high-frequency polling. The link outputs are browsed at a high frequency, and the ``DAQ'' words are read and concatenated into the wide output record. 
If $N$ inputs are used, this method requires the scanning frequency to be $N$ times higher than the data rate to avoid congestion at high “DAQ” data intensity.
This method has been successfully used for concentrating the data received from SMX2 frontend ASICs in a group of 14 E-Links in the CBM readout~\cite{cbm_collaboration_technical_2023}, where scanning at 160~MHz was sufficient.
 However, the required scanning frequency may be too high for more inputs or higher data rates.
The second method is based on width conversion in input channels. The ``DAQ'' data delivered in each channel are accumulated until they fill the output record.
The dedicated data multiplexer with round-robin priority finds and transmits the complete output record. 
This method significantly disturbs the time ordering of the concentrated data. The data from low-rate channels may get significantly delayed (waiting until they can fill the output record) compared to the data from high-rate channels.
The width converters must be implemented for each channel separately, resulting in increased resource consumption.
The practical usage of this method is not explicitly described in the literature. But as shown in~\cite{guminski_benes_2023},
 it may be found in publicly available firmware sources for the FELIX board used by the ATLAS experiment at CERN~\cite{wu_felix_2019}.
The third method has been introduced in~\cite{guminski_benes_2023} and assumes concentration by properly routing data to the desired position in the output record. The viable and tested implementation uses an interconnection network. Neither clock multiplication nor prolonged buffering of data is required. 
}

\added{The ``non-DAQ'' input data words should be removed before packing the data into the wide output words. However, in that case,
the output word must be assembled from the received ``DAQ'' words in the process of data concentration.
It is desired that the ``DAQ'' words are packed in the same time order as they are received from the links.}

\added{Before introducing the method proposed in~\cite{guminski_benes_2023} and further improved in this article, it is worth quickly reviewing alternative solutions. Their number is limited because, despite the high importance of efficient data concentration, it is very sparsely described in most papers about DAQ solutions.  
}

\added{Two alternative approaches are presented in~\cite{guminski_benes_2023}. The first one uses scanning of $N$ input links at a frequency $N$ times higher than the data rate. All data may be checked and written to the appropriate location in the output word, or discarded before the new data set arrives. That approach has been successfully used for the initial concentration of data in the CBM STS detector~\cite{cbm_collaboration_technical_2023}. However, it can't be used for high data rates and a large number of concentrated inputs.}

\added{The second method independently assembles the complete output records from the received ``DAQ'' data in each input channel. The complete records are then transmitted to the output with round-robin priority. 
In this method, data from low-rate channels may get significantly delayed (until their output record is filled) compared to the data from high-rate channels.
The practical usage of this method is not explicitly described in the literature. But as shown in~\cite{guminski_benes_2023},
 it may be found in publicly available firmware sources for the FELIX board used by the ATLAS experiment at CERN~\cite{wu_felix_2019}.}
 
\begin{figure}[htbp]
	\centering
	\includegraphics[width=.49\linewidth]{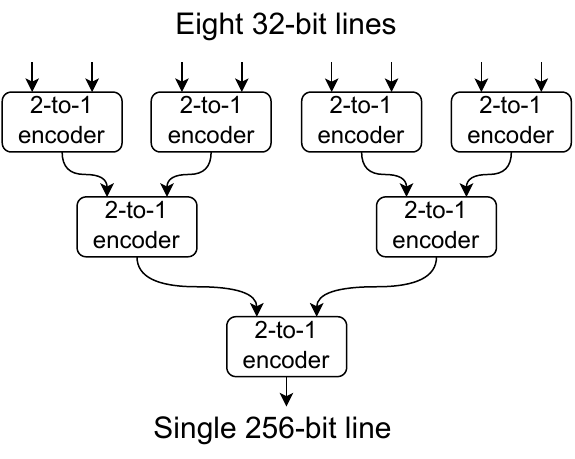}
	\caption{\label{fig:8-to-1-enc} 
	Structure of an 8-to-1 encoder built from 2-to-1 encoders. Figure reproduced (redrawn) and caption copied from~\cite{bassi_fpga-based_2023}: ``A FPGA-Based Architecture for Real-Time Cluster Finding in the LHCb Silicon Pixel Detector,'' by G. Bassi, L. Giambastiani, K. Hennessy, F. Lazzari, M. J. Morello, T. Pajero, A. Fernandez Prieto, G. Punzi, in IEEE Transactions on Nuclear Science, vol. 70, no. 6, pp. 1189-1201, June 2023, doi: 10.1109/TNS.2023.3273600, CC BY.
	}
\end{figure}

\begin{figure}[htbp]
	\centering
	\includegraphics[width=.69\linewidth]{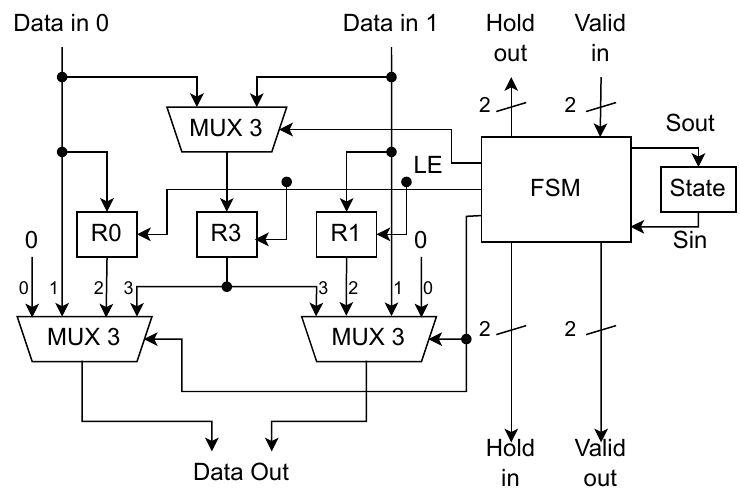}
	\caption{\label{fig:2-to-1-enc} 
	2-to-1 encoder block diagram. R0, R1, R3, and State are registers,
MUX0, MUX1, and MUX3 are multiplexers, and FSM is a finite state machine
that manages hold, valid, and LE write signals.
Figure reproduced (redrawn) and caption copied from~\cite{bassi_fpga-based_2023}: ``A FPGA-Based Architecture for Real-Time Cluster Finding in the LHCb Silicon Pixel Detector,'' by G. Bassi, L. Giambastiani, K. Hennessy, F. Lazzari, M. J. Morello, T. Pajero, A. Fernandez Prieto, G. Punzi, in IEEE Transactions on Nuclear Science, vol. 70, no. 6, pp. 1189-1201, June 2023, doi: 10.1109/TNS.2023.3273600, CC BY.
 	}
\end{figure}

\added{An interesting solution of concentrating eight 32-bit data streams into the 256-bit output has been developed for the LHCb Silicon Pixel Detector~\cite{bassi_fpga-based_2023}. It uses a three-level binary tree of blocks called ``2-to-1 encoders'', creating the final ``8-to-1 encoder'' (see Figure~\ref{fig:8-to-1-enc}). The data are concentrated first to four 64-bit words, then to two 128-bit words, and finally to a single 256-bit word.
Each ``2-to-1 encoder'' has a relatively complex internal structure, as shown in Figure~\ref{fig:2-to-1-enc}.
}

\added{A relatively complex logic was possible because the concentration worked at a low clock frequency 30~MHz.}

The third method has been introduced in~\cite{guminski_benes_2023} and assumes concentration by properly routing data to the desired position in the output record. 
\deleted{The viable and tested implementation uses an interconnection network. Neither clock multiplication nor prolonged buffering of data is required.}
\added{
In that method, neither multiplication nor prolonged buffering of data is required.
Such routing of data may be done with an interconnection network.
It is a very mature technology~\cite{noauthor_mathematical_1965}, widely used for data concentration in telecommunications and data processing systems~\cite{kumar_kundu_data_2014,Jain2020}.
However, the data concentration in those applications aims to transmit the data via reduced channels~\cite{jain_verifying_2016} and differs from the data concentration in DAQ described earlier.  Therefore, the problem studied in~\cite{guminski_benes_2023} and this paper is not a proposal of a new interconnection network  but adopting the well-known network architecture for that specific task and proving its correctness.
}

\subsection{Data concentration for DAQ with interconnection network}
\begin{figure}[htbp]
	\centering
	\includegraphics[width=.49\linewidth]{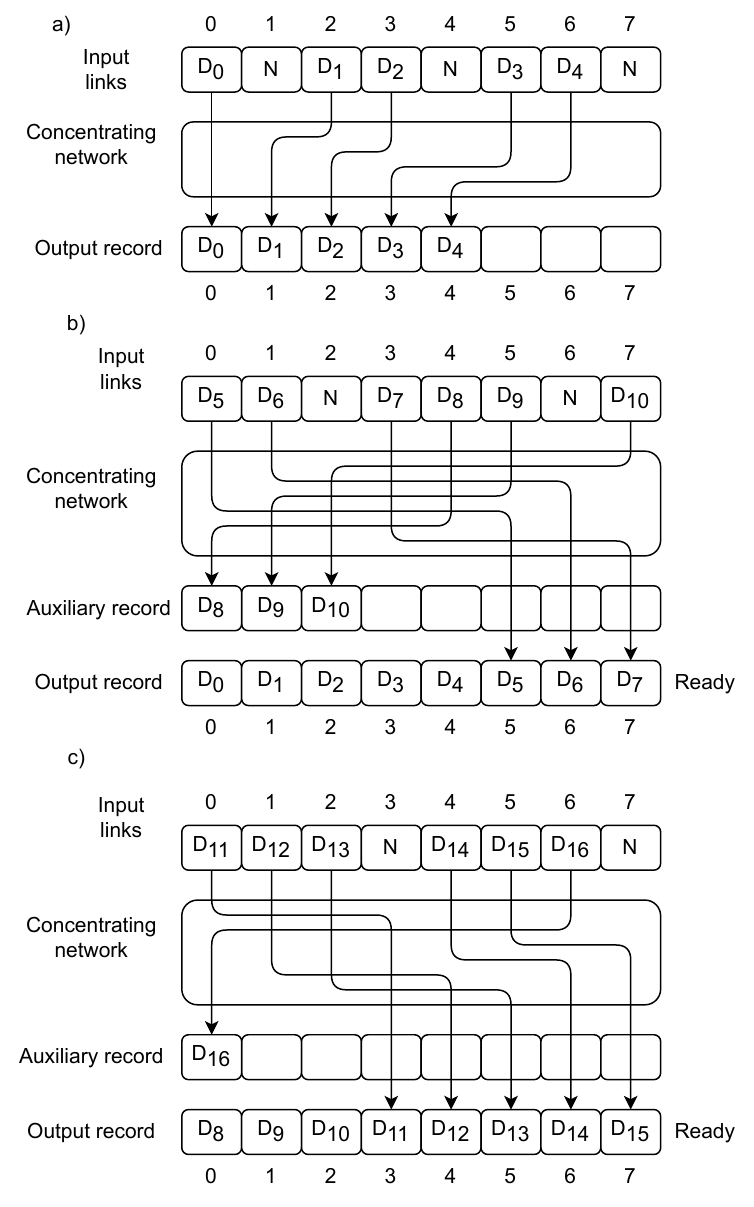}
	\caption{\label{fig:conc-ex1} Example of the concentration of data from 8 inputs to 8 outputs. The ``DAQ'' words are denoted by ``D'' with the number in subscript. The ``Non-DAQ'' words are denoted by ``N''. (a) In the first concentration cycle, five DAQ words are delivered via links. They are stored in the output record in locations 0 to 4. (b) In the second concentration cycle, the next 6 DAQ words are delivered. Three of them are stored in the output record in locations 5 to 7, but the next three must be stored elsewhere, as the locations 0 to 2 in the output record are still occupied. For that purpose, the auxiliary record is provided. The output record is completed and ready for sending to DAQ. (c) In the third concentration cycle, another six DAQ words are provided. Three DAQ words from the auxiliary record are copied to the output record. Therefore, filling the output records starts from location 3, and only 5 DAQ words may be stored. The sixth word must be stored again in the auxiliary record.
	}
\end{figure}

Example of data concentration with interconnection network is shown in Figure~\ref{fig:conc-ex1}.

\begin{figure}[htbp]
	\centering
	\includegraphics[width=.59\linewidth]{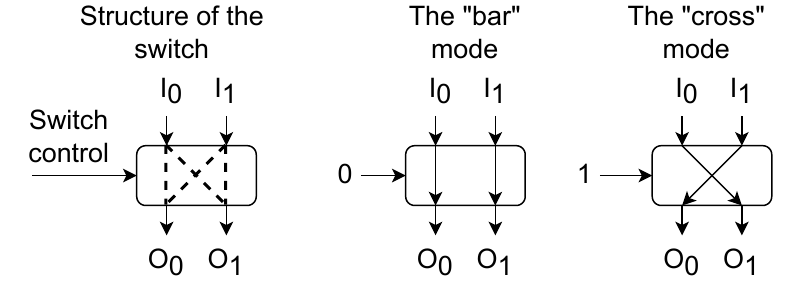}
	\caption{\label{fig:switch} Switch used in the concentrating network. Figure based on~\cite{chuan-lin_wu_class_1980}.
	}
\end{figure}

The data from the links are fed into the concentrating network, which removes the non-DAQ words and delivers the DAQ words to the consecutive locations in the output record. The output record may be partially filled at the end of the concentration cycle. Therefore, the location of the first DAQ word must be selectable.
Suppose after filling the output record, there are still DAQ words at the link outputs to be received. 
In that case, they must be stored in an additional auxiliary record, from which they are copied at the beginning of the next concentration cycle.
The concentrating network may be built as a multistage interconnection network~\cite{chuan-lin_wu_class_1980},
consisting of switches with two inputs and two outputs (see Figure~\ref{fig:switch}), transmitting the data transparently (``bar'' mode) or swapping them (``cross'' mode). 

To keep track of the current occupancy of the output and auxiliary records and to configure the concentrating network according to the availability of the DAQ words at inputs, an additional ``controller'' block is needed, which is not shown in Figure~\ref{fig:conc-ex1}.

\added{It counts the inputs delivering the DAQ words (``active inputs'') and assigns them the consecutive free positions in the output record. After the last position is assigned, the next active inputs are assigned the positions in the auxiliary record.}

\begin{figure}[htbp]
	\centering
	\includegraphics[width=.59\linewidth]{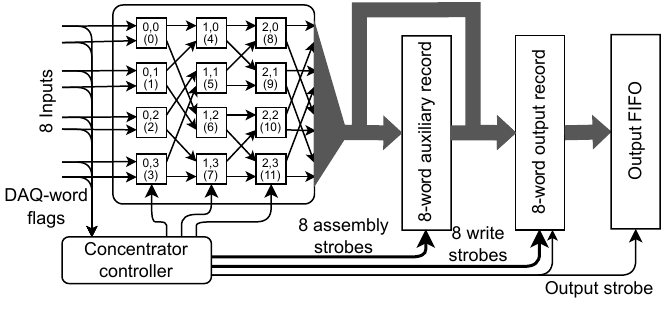}
	\caption{\label{fig:conc8} Concentration of data from 8 inputs. Slightly modified figure from~\cite{guminski_benes_2023}, according to CC BY license.
	}
\end{figure}

The development of a data concentrator with 8 inputs has been presented in~\cite{guminski_benes_2023}.
 Initially, the 8x8 Beneš network was used as the concentrating network, as it can perform any data permutation.
However, finding the proper configuration of switches for the particular permutation is a complex problem. Therefore, the brute force approach was used, and all needed configurations were stored in the table.
 During the development, it was found that the full Beneš network is not needed for data concentration. 
Therefore, finally, the concentrator based on an 8x8 ``reduced Beneš network'' has been proposed as the optimal solution (see Figure~\ref{fig:conc8}). 

However, concentrating data from more input streams is needed in certain applications.
An example may be a concentration of 32-bit words to a 512-bit output record (e.g., for 8-lane PCIe Gen 4 or 16-lane PCIe Gen 3)
or a concentration of 16-bit input words to a 256-bit output record.

Therefore, a concentrating network with a higher number of inputs and outputs is needed.

\section{Attempt to extend the previous concentrator}
\label{sec:old-approach}
The basis for the attempt to directly extend the previous concentrator was the assumption
that the ``reduced Beneš network'' (described in~\cite{guminski_benes_2023}) of any size could concentrate data.
However, without formal proof, this hypothesis could only be checked by explicitly testing all possible settings of switches.

For a 16x16 network, the number of switches is 32, so the number of possible configurations
 is ca. 4 billion. The tools developed in~\cite{guminski_benes_2023} have been appropriately extended, 
 and all possible configurations have been checked.
Indeed, it appeared that such a network can perform the concentration task. 
However, the problem is the considerable size of the table storing the necessary configurations of switches,
which must be used in the concentrator controller.
There are 32 switches to be configured, and the configuration depends on the ``DAQ word'' flags of 16 inputs
and the 4-bit number of occupied words in the output record.
Therefore, storing the configuration in the table requires the memory with $2^{16+4} = 2^{20}$ 32-bit words. 
Implementing it in the AMD/Xilinx FPGA would consume 1024 BRAMs or 128 UltraRAMs, which is unacceptable.

A solution could be transforming that table into a combinational function.
An attempt was made to use the generated configuration table as the truth table and perform the logic synthesis and optimization. The synthesis was attempted with two commercially available tools: AMD/Xilinx Vivado, Intel/Altera Quartus,
and the open-source Yosys~\cite{yosys-url} environment. 
Unfortunately, none of those attempts resulted in combinational logic of reasonable complexity.
\added{Vivado 2022.2 implemented it as a combinational logic, consuming 524189 LUTs in an Ultrascale+ FPGA.
Quartus 21.1 implemented it as a memory, consuming exactly $2^{20} \cdot 32 = 33554432$ memory bits.}
Therefore, this approach has been abandoned and replaced with the one based on the analysis of the concentration network.

\section{Analysis of the concentrating network}
\label{sec:analyzis}
The concentrating network described in~\cite{guminski_benes_2023} as the ``reduced Beneš network'' 
is, in fact, the ``baseline network''~\cite{chuan-lin_wu_class_1980} with a bit-reversed order of outputs.
Further, it will be described as a ``baseline network with reversed outputs'', or BNRO in short.

\begin{figure}[htbp]
	\centering
	\includegraphics[width=.49\linewidth]{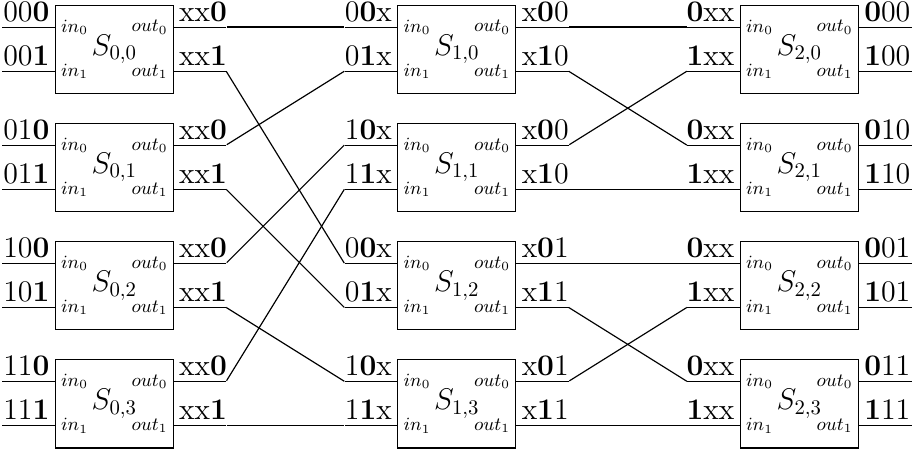}
	\caption{\label{fig:bnet3} Concentration of data from 8 inputs to 8 outputs.
		The binary numbers at the inputs of the switches describe from which network inputs the particular input may receive data.
		The binary numbers at the outputs of the switches describe to which network outputs the particular output may deliver data. The digits in bold are used to find the required mode of the specific switch.
	}
\end{figure}

\begin{figure}[htbp]
	\centering
	\includegraphics[width=.79\linewidth]{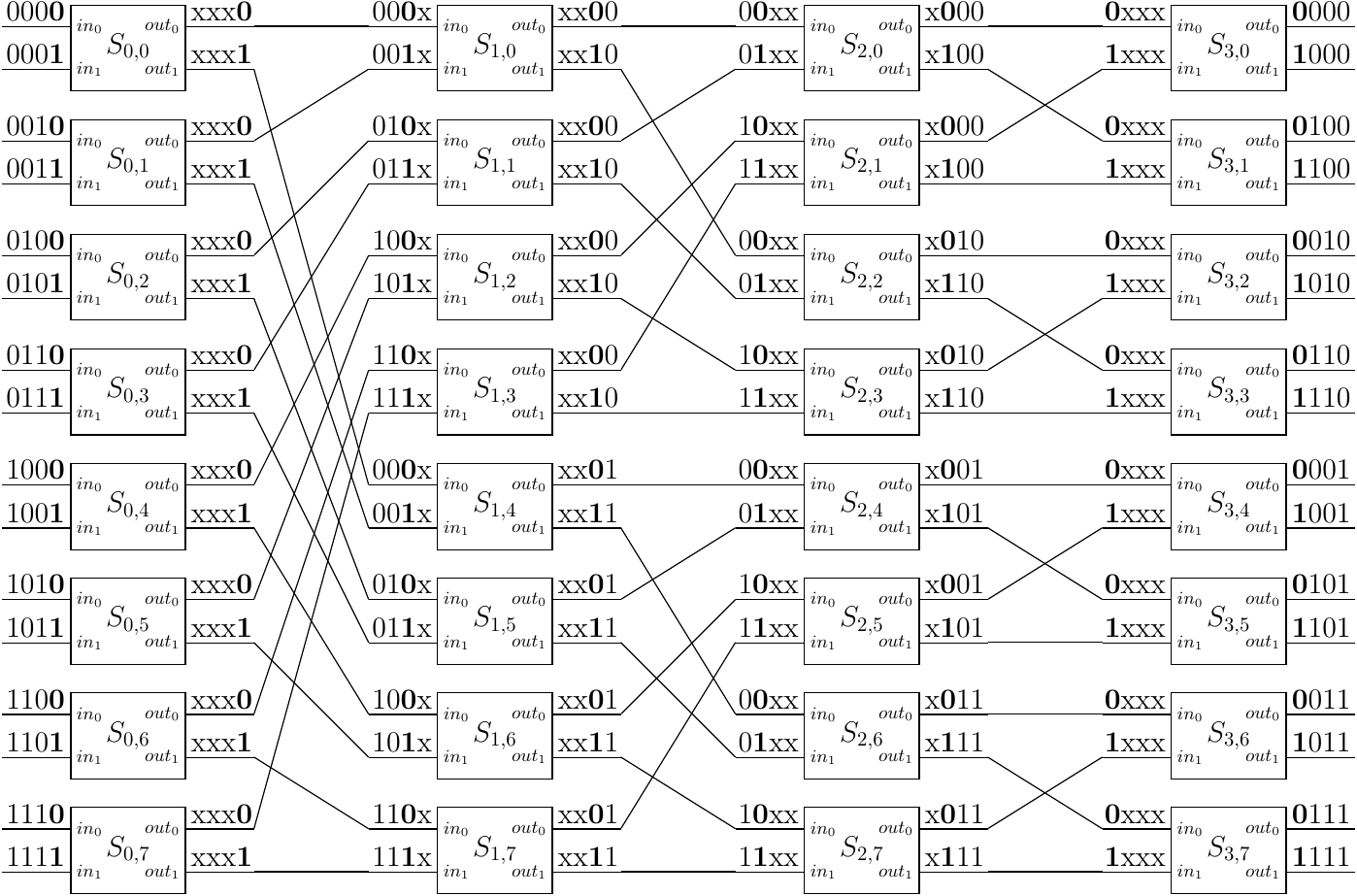}
	\caption{\label{fig:bnet4} Concentration of data from 16 inputs to 16 outputs.
		The binary numbers at the inputs of the switches describe from which network inputs the particular input may receive data.
		The binary numbers at the outputs of the switches describe to which network outputs the particular output may deliver data. The digits in bold are used to find the required mode of the specific switch.
	}
\end{figure}

\begin{figure}[htbp]
	\centering
	\includegraphics[width=.99\linewidth]{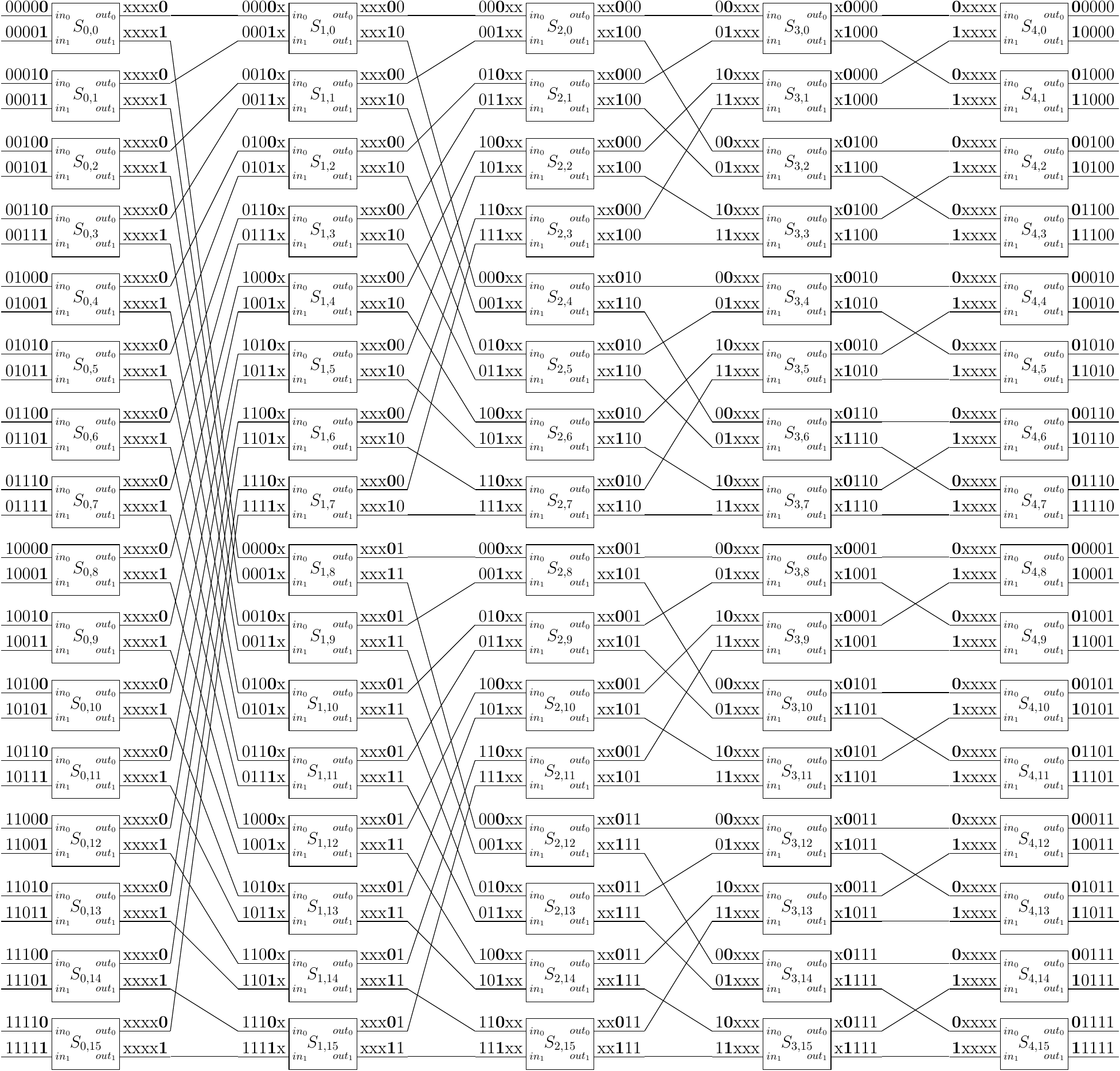}
	\caption{\label{fig:bnet5} Concentration of data from 32 inputs to 32 outputs.
		The binary numbers at the inputs of the switches describe from which network inputs the particular input may receive data.
		The binary numbers at the outputs of the switches describe to which network outputs the particular output may deliver data. The digits in bold are used to find the required mode of the specific switch.
	}
\end{figure}

The topology of this network for different numbers of layers (and what is related -- different numbers of inputs and outputs) 
is shown in Figures~\ref{fig:bnet3} to~\ref{fig:bnet5}.

\begin{figure}[htbp]
	\centering
	\includegraphics[width=.69\linewidth]{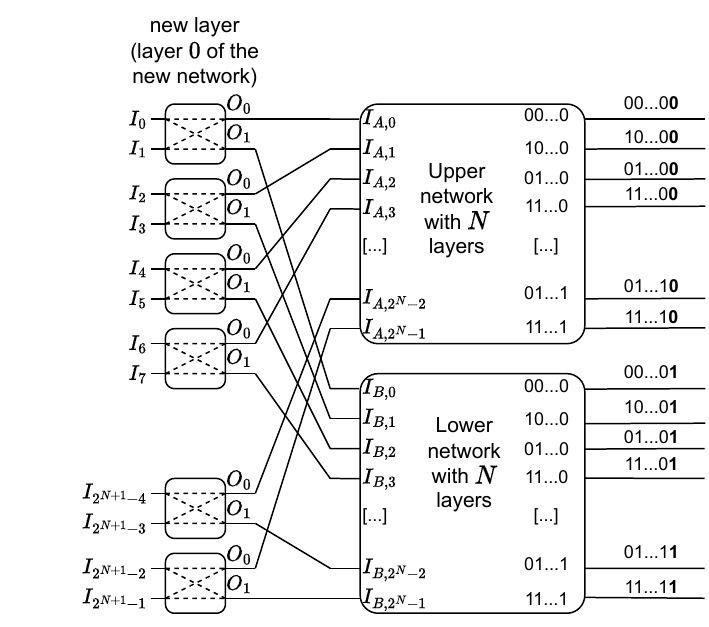}
	\caption{\label{fig:baseline-recursive} 
		Recursive construction of the BNRO network with $N+1$ layers from two smaller BNRO networks with $N$ layers.
		The numbers of outputs are in binary form to show that the switch output number in the added layer 
		determines the $0^\mathrm{th}$ bit in the output number of the new network.
		Diagram based on~\cite{chuan-lin_wu_class_1980}.
	}
\end{figure}

The structure of the baseline network is regular. The network with $N+1$ layers may be created from two baseline networks with $N$ layers
by adding a layer with switches where output $0$ is connected to the ``upper network'' and output $1$ is connected to the ``lower network'' (see Figure~\ref{fig:baseline-recursive}).

The rule for determining the output switch also defines the simple and unambiguous routing algorithm:
To route the data through the $N$-layer BNRO network from input $k$ to output $m$, we:
\begin{itemize}
	\item Start from the input $k$ of the network,
	\item Compare the bit $0$ in the input number with the bit $0$ of the target output number. The switch should be set to the ``bar'' mode if the bits are equal. If they differ, the switch should be set to the ``cross'' mode.  After setting the switch in the suitable mode, find the right switch and its input in the $1^\textrm{st}$ layer.
	\item Repeat the next steps in the loop until the output of the network is reached:
	\begin{itemize}
		\item In layer $l$, compare the $l^\textrm{th}$ bit in the target output and the network input numbers.
		 The switch should be set to the ``bar'' mode if the bits are equal. If they differ, the switch should be set to the ``cross'' mode.  After setting the switch in the suitable mode, find the right switch and its input in the next layer.
		\item Increase $l$.		
	\end{itemize}
\end{itemize}

The above algorithm enables routing any input of the BNRO network to any of its outputs.
However, it does not warrant that the BNRO network can perform any possible data permutation.
The routing is impossible if the data delivered to two inputs of a particular switch should be routed to the same output of that switch,
 which results in a collision.

The results obtained in~\cite{guminski_benes_2023} 
and~\added{experiments with 16x16 (4-layer) network in} Section~\ref{sec:old-approach} 
suggest that such collision should not happen in the BNRO network 
when used for data concentration. However, formal proof is needed.

\section{Proof of the BNRO suitability for data concentration}
The suitability of the BNRO network to data concentration may be proven in various ways. Below, two of them are presented.

\subsection{Proof based on the mathematical induction}
\label{sec:ind-proof}
\begin{figure}[htbp]
	\centering
	\includegraphics[width=.69\linewidth]{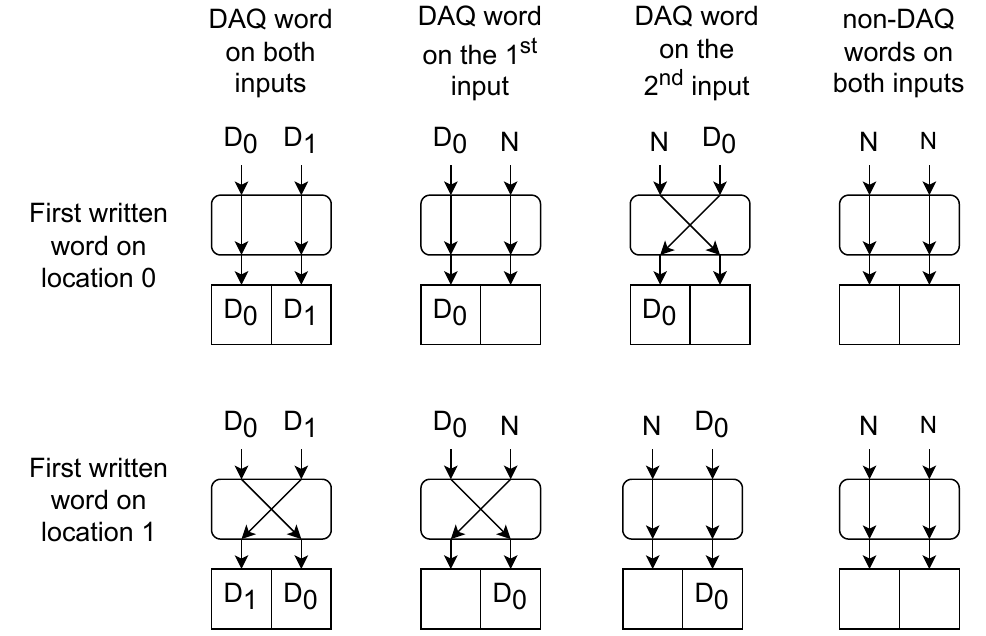}
	\caption{\label{fig:bnro-1lay-demo} All possible cases of data concentration in 1-layer BNRO.
	}
\end{figure}

We can take the network with one layer ($N=1$) and check that it can perform the concentration task. That is trivial for two inputs and outputs - the network can propagate the data without swapping them (corresponding to writing the output record from location 0) or swapping them (corresponding to writing the output record from location 1).
 The corresponding data may be simply ignored if one or both inputs deliver a non-DAQ word.
  So, BNRO with $N=1$ layer can concentrate data  (see Figure~\ref{fig:bnro-1lay-demo}).

Then, we must prove that if a BNRO network with $N$ layers can concentrate data, the network with $N+1$ layers, created as shown in Figure~\ref{fig:baseline-recursive}, can also do it.

Let us assume that the first DAQ word should be routed to the output $t < 2^{n+1}$. If the $N+1$ network receives the data with 
$m < 2^{n+1}$ DAQ words, that set of words may be split into two subsets with sizes differing by no more than 1.
If $t$ is even, we route the first DAQ word to the upper subnetwork. If $t$ is odd, we route that word to the lower subnetwork.
The next DAQ words (if available) must be routed alternately to both subnetworks.
 Such a routing is always possible. 
 
 Suppose the currently handled DAQ word is connected to the switch with another input receiving the non-DAQ word. In that case, we can select the output and the upper or lower subnetwork as required.
 
 Suppose the currently handled DAQ word is connected to the switch where another input also receives the DAQ word. In that case, we may route it to the required network, and the next DAQ word will be automatically routed to the opposite network, fulfilling the requirement of the alternate routing.

As each subnetwork can perform the data concentration, both the even and odd DAQ words may be routed to the appropriate subnetworks' consecutive (in modulo $2^N$ sense) outputs.

\begin{figure}[htbp]
	\centering
	\includegraphics[width=.79\linewidth]{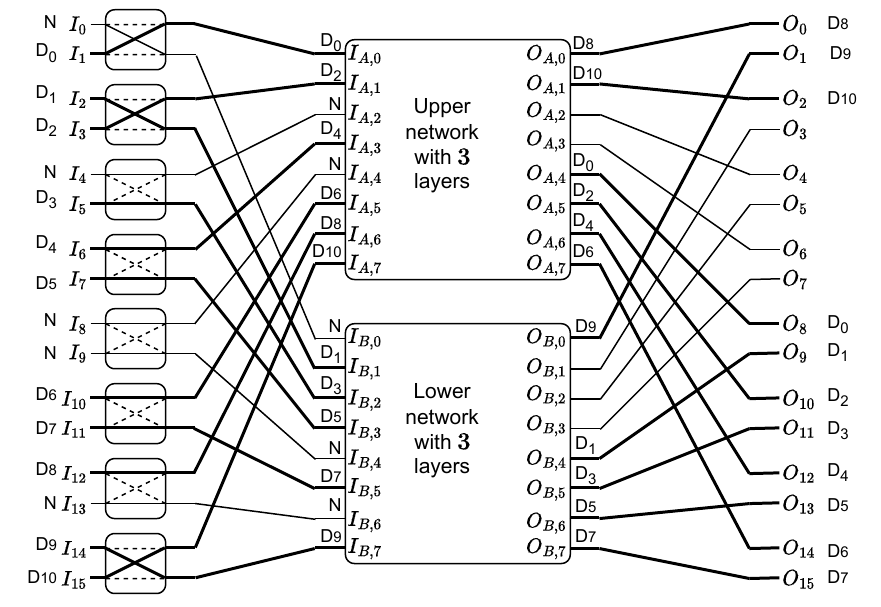}
	\caption{\label{fig:recursive-routing-even} Example of concentration of data in 4-layer BNRO assuming that 3-layer BNRO can concentrate a subset of data. The case with an even number of the first location in the output record ($t=8$) is shown. As described in section~\ref{sec:ind-proof}, the even data are passed to the upper subnetwork and routed starting from the output $t/2=4$. The odd data are passed to the lower subnetwork and routed starting from output $t/2=4$. After merging the outputs, the data are correctly concentrated, starting from output $t=8$. Please note how the output numbers have wrapped around. Data $D_8$, $D_9$, and $D_{10}$ are delivered to outputs 0, 1 and 2. For clarity, the outputs of the subnetworks are shown in the natural order (not in reversed order).
	}
\end{figure}

If $t$ is even, we configure the upper subnetwork so that even DAQ words are routed to its outputs starting from the $t/2$ output and the lower subnetwork so that the odd DAQ words are routed to its outputs starting from the $t/2$ output. In the whole network, those data will appear at consecutive (in modulo $2^N$ sense) outputs starting from output $t$. An example of such concentration for 4-layer BNRO is shown in Figure~\ref{fig:recursive-routing-even}.

\begin{figure}[htbp]
	\centering
	\includegraphics[width=.79\linewidth]{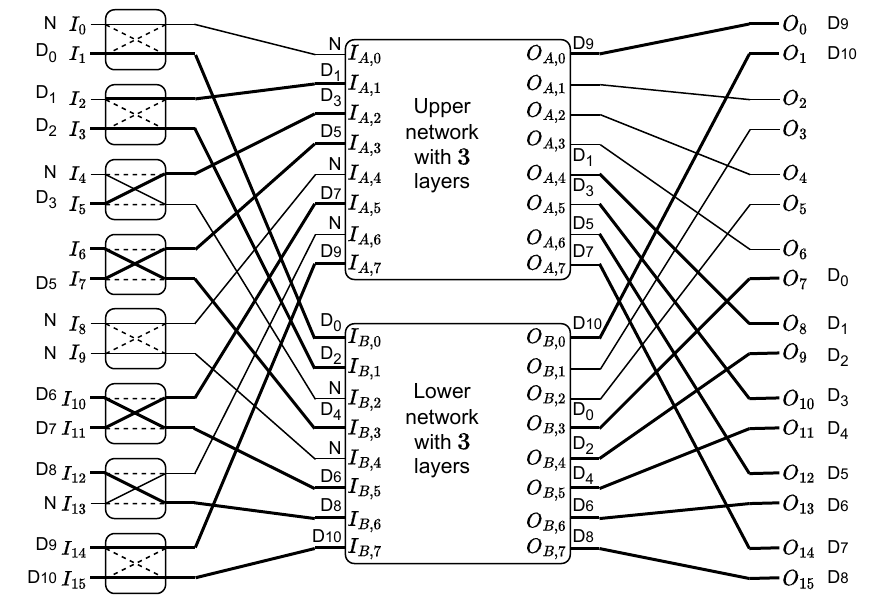}
 	\caption{\label{fig:recursive-routing-odd} Example of concentration of data in 4-layer BNRO assuming that 3-layer BNRO can concentrate a subset of data. The case with an odd number of the first location in the output record ($t=7$) is shown. As described in section~\ref{sec:ind-proof}, the even data are passed to the upper subnetwork and routed starting from the output $(t-1)/2=3$. The odd data are passed to the lower subnetwork and routed starting from the output $(t+1)/2=4$. After merging the outputs, the data are correctly concentrated, starting from the output $t=7$. Please note how the output numbers have wrapped around. Data $D_9$ and $D_{10}$ are delivered to outputs 0 and 1. For clarity, the outputs of the subnetworks are shown in the natural order (not in reversed order).			
	}
\end{figure}

If $t$ is odd, we configure the lower subnetwork so that even DAQ words are routed to its outputs starting from $(t-1)/2$ output and the upper subnetwork so that the odd DAQ words are routed to its outputs starting from $(t+1)/2$ output. Those data will appear at consecutive (in modulo $2^N$ sense) outputs starting from ouput $t$ in the whole network. An example of such concentration for 4-layer BNRO is shown in Figure~\ref{fig:recursive-routing-odd}.

Based on the mathematical induction principle, the above proves that BNRO with any number of layers $N \ge 1$ can concentrate data.

\subsection{Proof based on analysis of collision possibility}

\begin{figure}[htbp]
	\centering
	\includegraphics[width=.59\linewidth]{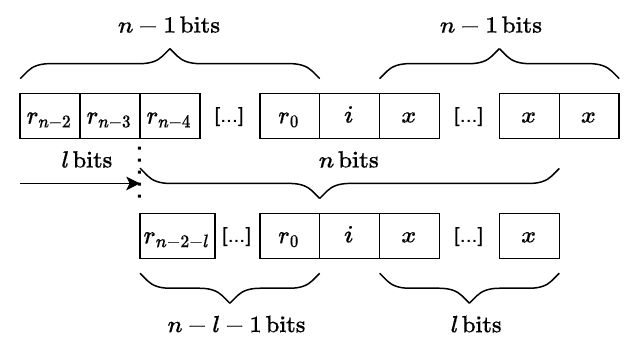}
	\caption{\label{fig:bnet-inputs} Diagram showing the dependency between the number of layer $l$ and the number of switch $r$,
		 and the numbers of the network inputs from which the data may be routed to the input $i$ of that switch. 
		 The bits creating the numbers of the accessible network inputs (the lower row) are selected from the vector consisting of bits of the number of the switch ($r$), the bit corresponding to the number of the switch input ($i$), and variable bits $x$ depending on the configuration of the switches in the preceding layers (the upper row).
	}
\end{figure}

\begin{figure}[htbp]
	\centering
	\includegraphics[width=.59\linewidth]{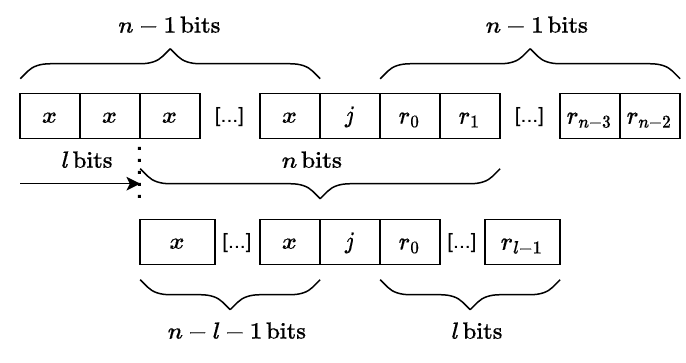}
	\caption{\label{fig:bnet-outputs} Diagram showing the dependency between the number of layer $l$ and the number of switch $r$,
		 and the numbers of the network outputs to which the data may be routed from the input $j$ of that switch.	
		 The bits creating the number of the accessible network outputs (the lower row) are selected from the vector (the upper row) consisting of variable bits $x$ depending on the configuration of the switches in the following layers, the bit corresponding to the number of the switch output ($j$) and bits of the number of the switch ($r$).	
	}
\end{figure}

Analyzing the recursive building of the BNRO network and analyzing topologies
of BNRO for small numbers of layers
 (see Figures~\ref{fig:bnet3},\ref{fig:bnet4} and \ref{fig:bnet5}), 
we may find the rules describing the numbers of network inputs that
 may be connected
to the particular input $i$ of switch number $r$ located in layer $l$, and numbers of network outputs that may be connected to the particular output $j$ of that switch.
These rules are presented in Figures~\ref{fig:bnet-inputs} and~\ref{fig:bnet-outputs}.

\begin{figure}[htbp]
	\centering
	\includegraphics[width=.99\linewidth]{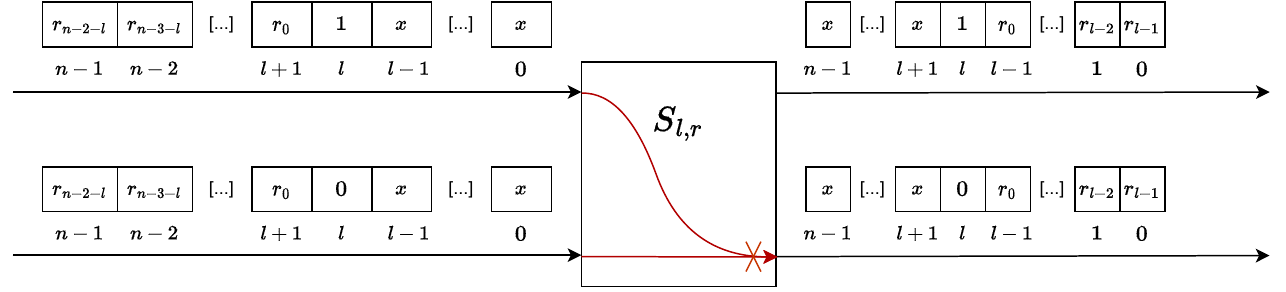}
	\caption{\label{fig:collision} Diagram showing a single switch with a collision in data routing. The collision occurs when the data delivered to both switch inputs should be routed to the same output.
	}
\end{figure}

With the above rules, we may analyze the conditions resulting in impossible data routing due to a collision in a certain switch.
The collision happens when data from both inputs of the particular switch should be routed to the same output, as shown in Figure~\ref{fig:collision}.

\begin{figure}[htbp]
	\centering
	\includegraphics[width=.99\linewidth]{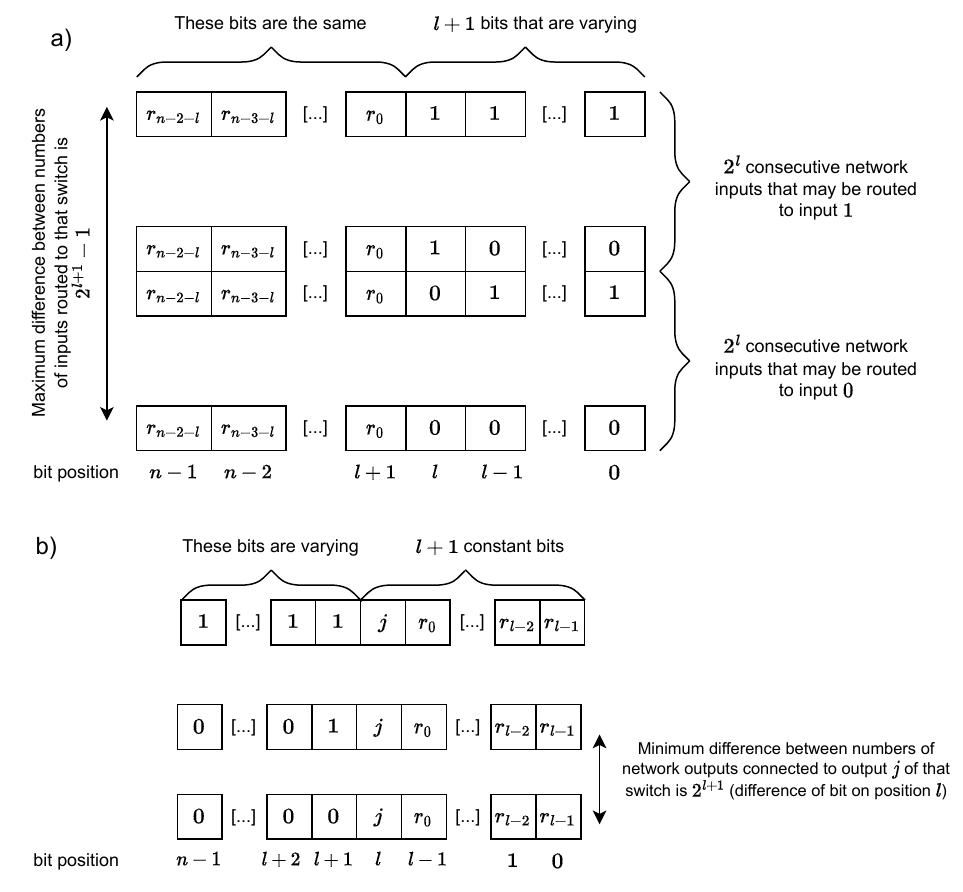}
	\caption{\label{fig:in-out-numbers} 
		Diagram showing numbers of the network inputs from which data may be routed to the inputs of a particular switch and numbers of network outputs to which the data from its specific output may be routed.
		(a) The numbers of network inputs connected to the switch in layer $l$ may differ by no more than $2^{l+1}-1$. 
		(b) The numbers of network outputs connected to that switch must differ at least by $2^{l+1}$.
	}
\end{figure}

The network inputs accessible from the individual input have the $2^l$ consecutive numbers. The group of numbers associated with inputs $0$ and $1$ are neighboring, so all inputs accessible from the switch have $2^{l+1}$ consecutive numbers (see Figure~\ref{fig:in-out-numbers} (a)). Therefore, the difference between the numbers of two associated inputs must be not higher than $2^{l+1}$.

The outputs accessible from the switch have numbers spread equally and differ by $N \cdot 2^{l+1}$.
 Therefore, the difference between the numbers of two accessible outputs must be not less than $2^{l+1}$ (see Figure~\ref{fig:in-out-numbers} (b)).

Suppose the collision in the switch should happen. In that case, the data from two network inputs accessible from the switch (with numbers differing by not more than $2^{l+1}-1$) should be routed to different network outputs accessible from the same switch output.
However, the numbers of those network outputs must differ by not less than $2^{l+1}$.
The concentration of data, however, does not insert new data words into the data stream. It may only remove certain not needed data words from that stream. Therefore, during the concentration, for each pair of data words processed in the same concentration cycle, the difference of numbers between the destination network outputs must be not higher than the difference between the numbers of their network inputs.

That means the collision never happens when the analyzed network concentrates the data, and the BNRO network can perform the concentration task.

\section{Implementation}
\begin{figure}[htbp]
	\centering
	\includegraphics[width=.69\linewidth]{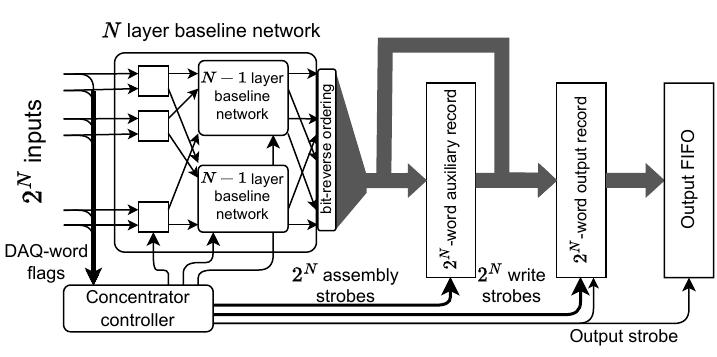}
	\caption{\label{fig:conc-N} The block diagram of the scalable data concentrator based on an $N$-layer baseline network with reversed outputs (BNRO). It is an extended and scalable version of the concentrator shown in Figure~\ref{fig:conc8}.
	}
\end{figure}

The concentrator based on the BNRO network has been implemented in the VHDL language. Its block diagram is presented in Figure~\ref{fig:conc-N}.

Its sources are publicly available under the dual GPL/BSD license in the Gitlab repository~\cite{url-xconcentrator-gitlab}.

The sources are highly parameterized. The user may define the number of layers in the concentrating network and, therefore, the number of inputs and outputs. It is also possible to define the type of data that are concentrated.

The repository contains the testbench, allowing the user to verify the correct operation of the concentrator. As the payload, the integer values are used. The DAQ words with consecutive integer values are delivered to the consecutive inputs. 
The user may set the probability of the data being accepted by an input. If the data is not accepted, the input is skipped, and the data is tried to be delivered to the next input. 
The data leaving the concentrator are written to the file. If the concentrator works correctly, the output file contains the consecutive integers. A dedicated Python script checks the content of the generated file.

\begin{figure}[htbp]
	\centering
	\includegraphics[width=.99\linewidth]{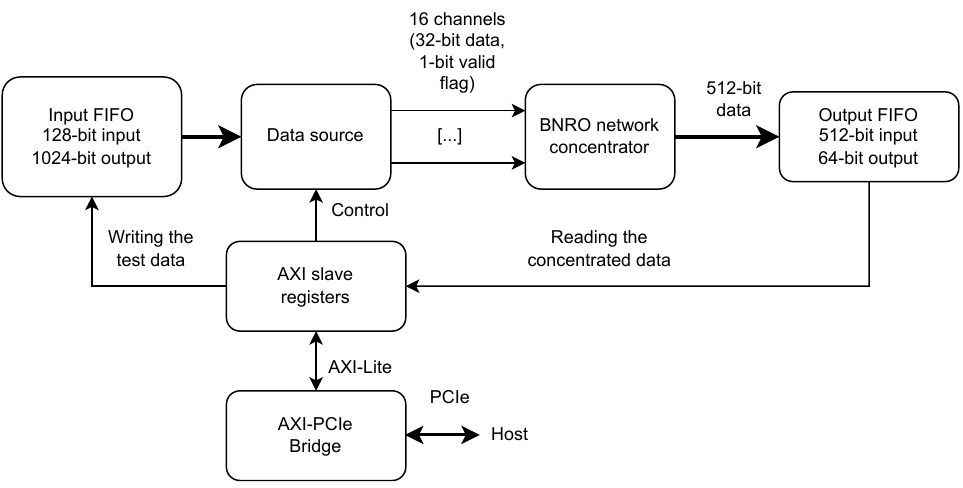}
	\caption{\label{fig:test_hw} Testbench for testing the concentrator in the hardware. Slightly modified figure from~\cite{guminski_benes_2023}, according to CC BY license.
	}
\end{figure}

Additionally, hardware testbenches for the KCU105~\cite{url_xlx_kcu105} and KCU116~\cite{url_xlx_kcu116} boards were created for the 16-input concentrator.
 Similarly, like in~\cite{guminski_benes_2023}, the setup contains 
FIFOs for storing the input data to be fed into the concentrator and its output data.
The block diagram of the test setup is shown in Figure~\ref{fig:test_hw}. 

The testbench is controlled with a Python script using the standard \textbf{uio\_pci\_generic} driver.
Both the script and the testbench have been extended to support a higher number of concentrated inputs.

\section{Tests and results}
The simulation tests were performed for 4 layers (16 inputs) and for 5 layers (32 inputs).
Configurations without pipeline registers, with pipeline registers in all layers, and with pipeline registers in selected layers were simulated.
Simulations were repeated with different values of the probability of input data availability.
All simulations have confirmed that the concentrator works correctly, and the concentrated words are delivered in the correct order as a dense data stream without ``holes''.

For tests in the hardware, the synthesis and implementation with Vivado 2021.2 were performed for two variants -- with pipeline registers inside switches and without them.
For the KCU116 board, both variants obtained the correct timing closure for 250~MHz clock frequency.
The resource consumption is summarized in Table~\ref{tab:synthesis-results-kcu116}. 

\begin{table}[htbp]
	\caption{Resource consumption of the 16-input BNRO-based data concentrator together with the testbench for the KCU116 platform.
		Absolute and percentage (in parenthesis) consumption is given.
		Separate values for the testbench and concentrator itself are given. \added{The ``non-registered switches'' configuration does not use pipeline registers in the concentrating network. The ``registered switches configuration'' uses pipeline registers in switches in all layers. That increases the resource consumption and latency but may increase the maximum clock frequency.}
		\label{tab:synthesis-results-kcu116}}
	\newcolumntype{C}{>{\centering\arraybackslash}X}
	\begin{tabularx}{\textwidth}{CCCCCCC}
		\hline
		~ &\multicolumn{3}{c}{\textbf{With non-registered switches}} & \multicolumn{3}{c}{\textbf{With registered switches}} \\
		~ &\textbf{LUTs}	& \textbf{Flip Flops}	& \textbf{Block RAMs} &\textbf{LUTs}	& \textbf{Flip Flops}	& \textbf{Block RAMs}\\
		\hline
		\textbf{Available}	& 216960 & 433920 & 480	& 216960 & 433920 & 480\\
		\textbf{Whole testbench} 
		& 17741 (8.17\%) & 19921 (4.59\%) &  69 (14.37\%) 
		& 17767 (8.19\%) & 22113 (5.10\%) &  69 (14.37\%) \\
		\textbf{Data concentrator}
		& 3380 (1.56\%) & 1640 (0.38\%) & 0 (0.0\%) 
		& 3410 (1.57\%) & 3830 (0.88\%) & 0 (0.0\%) \\			  
		\hline
	\end{tabularx}
\end{table}

 For KCU105 board, only the version with pipeline registers could work with a 250~MHz clock. The timing closure could be obtained for clock frequency below 197~MHz in the version without pipeline registers. \added{Without the pipeline registers, the whole concentrating network is implemented as a combinational logic, which results in long critical path. Adding pipeline registers in switches in all layers or in selected ones shortens the combinational path reducing the propagation time between registers and thence increasing the maximum clock frequecy.}
The resource consumption is summarized in Table~\ref{tab:synthesis-results-kcu105}.

\begin{table}[htbp]
	\caption{Resource consumption of the 16-input BNRO-based data concentrator together with the testbench for the KCU105 platform.
		Absolute and percentage (in parenthesis) consumption is given.
		Separate values for the testbench and concentrator itself are given. \added{The ``non-registered switches'' configuration does not use pipeline registers in the concentrating network. The ``registered switches configuration'' uses pipeline registers in switches in all layers. That increases the resource consumption and latency but may increase the maximum clock frequency.}
		\label{tab:synthesis-results-kcu105}}
		\newcolumntype{C}{>{\centering\arraybackslash}X}
		\begin{tabularx}{\textwidth}{CCCCCCC}
			\hline
			~ &\multicolumn{3}{c}{\textbf{With non-registered switches}} & \multicolumn{3}{c}{\textbf{With registered switches}} \\
			~ &\textbf{LUTs}	& \textbf{Flip Flops}	& \textbf{Block RAMs} &\textbf{LUTs}	& \textbf{Flip Flops}	& \textbf{Block RAMs}\\
			\hline
			\textbf{Available}	& 242400 & 484800 & 600	& 242400 & 484800 & 600\\
			\textbf{Whole testbench} 
			& 9279 (3.83\%) & 10342 (2.13\%) &  50 (8.33\%) 
			& 9387 (3.87\%) & 11619 (2.40\%) &  50 (8.33\%) \\
			\textbf{Data concentrator}
			& 3273 (1.35\%) & 2714 (0.56\%) & 0 (0.0\%) 
			& 3436 (1.42\%) & 3921 (0.81\%) &  0 (0.0\%) \\			  
			\hline
		\end{tabularx}
\end{table}

The three configurations capable of operating at 250~MHz frequency were tested in the FPGA boards inserted into PCIe Gen3x8 slots in a Linux PC.
The testbench was controlled with the prepared Python script.
As in simulations, the tests were repeated with different values of the probability of input data availability.
All the tests have confirmed the correct operation of the concentrator.

\section{Discussion}
The data concentration concept presented in~\cite{guminski_benes_2023} significantly improved the concentration performance compared to the previously used methods like high-frequency polling or width conversion in the input channels.
It eliminated the necessity to use very high clock frequencies and disturbances in the time ordering of the concentrated data.

The disadvantage of that solution was poor scalability.
 It was based on an 8x8 interconnection network. 
 Therefore, natively, it supported up to 8 inputs and the output word with a length eight times bigger than the input words.

   Specific extensions have been proposed in~\cite{guminski_benes_2023} increasing the number of inputs to 9 or 12 but at the cost of introducing additional clocks with slightly higher frequency and higher design complexity.
Attempt to use a bigger 16x16 network was blocked by problems with too large a lookup table storing the configuration of switches in the interconnection network.

The new solution described in this paper extends this concept with scalability. It proves that the $N$-layer (with $2^N$ inputs and outputs) baseline network with reversed outputs (BNRO) can correctly concentrate the data. It significantly simplifies the design of such a network, offering an efficient and synthesizable implementation of the BNRO controller.
The new method enabled easy implementation of a concentrator with 4 layers (16 inputs), confirmed in hardware, and with 5 layers (32 inputs), confirmed in simulations. 
\deleted{
Theoretically, even a higher number of layers may be used. 
The important feature of the proposed architecture is the fact that it may be pipelined. Increasing the number of layers may increase the concentration latency but does not increase the critical path length in the concentrating network.}

\deleted{However, the increased number of inputs affects the complexity of the network controller. The algorithm for finding the necessary configuration of switches (see Section~\ref{sec:analyzis}) may be pipelined because the switch settings must be delivered to each layer only with the data.  
The sensitive part of the current solution is the algorithm that counts the number of inputs delivering the ``DAQ'' data, calculates the target output for each data, and the first written position in the next concentration cycle.
These tasks must be performed in one concentration clock period. The critical path length linearly depends on the number of inputs in the current implementation. That may limit the speed at a higher number of inputs. Future research should be oriented toward the optimization of that algorithm.}
\added{
Currently, the maximum 512-bit size of the output word seems reasonable. Even the planned solutions for PCIe Gen5 assume that width~\cite{url_xlx_versal_pcie}. With the 16-bit minimum width of concentrated data (it must contain the payload and the source ID), it gives the number of concentrated inputs equal to 32.}

\added{In the concentrating network, nothing prevents adding the $6^{\textrm{th}}$ layer and increasing the number of inputs to 64 if the width of the output word rises to 1024 in the future.}

\added{However, the number of inputs may affect the critical path in the network controller. In the current design, where to route the data from the particular input is elaborated in a single clock cycle. 
That operation includes counting the active inputs (those delivering the DAQ words). Then, the target output is calculated based on the active input number and the current occupancy of the output record (see Section~\ref{sec:analyzis}).
That results in roughly linear dependency of the critical path in the controller on the number of inputs.
However, the input stage of the concentrator may also be pipelined at the cost of additional latency. The input data may propagate through a required number of pipeline registers. That enables counting the active inputs and calculating the target outputs in a few cycles. Finding the optimal solution should be the object of future research.
}

\section{Conclusions}
The paper significantly improves and extends the data concentration method proposed in ~\cite{guminski_benes_2023}.
It proves that the baseline network with reversed outputs (BNRO) with any number of layers may concentrate data without the risk of collision.
An efficient algorithm for finding the necessary configuration of switches has been formulated, eliminating the need for huge configuration tables. 
The reference implementation of the concentrator, including the BNRO and its controller, has been created as a parameterized VHDL code.
Introducing optional pipeline registers in individual BNRO layers allows the user to find the optimal compromise between resource consumption, concentration latency, and maximum clock frequency.
The design has been tested in simulations in configurations with 4 and 5 layers.
The 4-layer version has been implemented in two different FPGA boards and successfully tested in the actual hardware.
\added{The confirmed achieved throughput for 16 32-bit inputs at 250~MHz was 128~Gb/s.}

The proposed concentrator may be a good solution for systems where high-speed and low-latency concentration of multiple short-word data streams into a long-word output is necessary.

Thanks to the availability of sources in the public repository under a permissive double GPL/BSD license, the solution may be widely reused and modified for particular needs.

\funding{The work has been partially supported by the statutory funds of the Institute of Electronic Systems. This project has also received funding from the European Union’s Horizon 2020 research and innovation programme under grant agreement No 871072.}

\dataavailability{The source code of the described concentrator is available under a free and open dual GPL/BSD licence in the public repository~\cite{url-xconcentrator-gitlab}.} 




\conflictsofinterest{The author declares no conflict of interest.} 




\begin{adjustwidth}{-\extralength}{0cm}

\reftitle{References}


\bibliography{xconcentrator}
\end{adjustwidth}
\end{document}